# Effect of the wire width on the intrinsic detection efficiency of superconducting-nanowire single-photon detectors

R. Lusche,[*] A. Semenov, H.-W. Hübers
*DLR Institute of Planetary Research, Rutherfordstrasse 2, 12489 Berlin, Germany*

K. Ilin, M. Siegel
*Institute of Micro- und Nano-electronic Systems (IMS), Karlsruhe Institute of Technology (KIT), Hertzstrasse 16, 76187 Karlsruhe, Germany*

Y. Korneeva, A. Trifonov, A. Korneev, G. Gol'tsman
*Moscow State Pedagogical University, 01069 Moscow, Russia*

D. Vodolazov
*Institute for Physics of Microstructures, Russian Academy of Sciences, 603950, Nizhny Novgorod, GSP-105, Russia*

Thorough spectral study of the intrinsic single-photon detection efficiency in superconducting TaN and NbN nanowires with different widths shows that the experimental cut-off in the efficiency at near-infrared wavelengths is most likely caused by the local deficiency of Cooper pairs available for current transport. For both materials the reciprocal cut-off wavelength scales with the wire width whereas the scaling factor quantitatively agrees with the hot-spot detection models. Comparison of the experimental data with vortex-assisted detection scenarios shows that these models predict a stronger dependence of the cut-off wavelength on the wire width.

## I. INTRODUCTION

Recently a noticeable progress has been achieved in understanding the mechanisms of single-photon detection in current-carrying superconducting nanowires that greatly extends the initial simplified hot-spot model.[1,2] The hot-spot detection scenario predicts a sharp cut-off in the detection efficiency that should occur at a certain point when the energy of the incoming photons decreases. However, even the refined hot-spot model[3] failed to explain the experimentally found gradual decay of the detection efficiency beyond the cut-off. This observation brought another, vortex-assisted detection scenario.[4,5,6] This quasistatic approach considers circularly symmetric magnetic vortices driven by the Lorentz force, which cross the entire width of the wire over the potential barrier. In this model the shape of the barrier and, consequently, the detection efficiency is extremely sensitive to the energy of the vortex core and to the details of the vortex nucleation at small distances to the wire edge.[7,8,9] For each photon energy there is a current saturating the local rate of vortex crossing at the absorption site.[6] Hence, for a particular current the wavelength dependence of the count rate should exhibit a kink which one may associate with the experimentally observed spectral cut-off. Another approach invokes the numerical solution of the time-dependent Ginsburg-Landau equations coupled with the two-dimensional heat diffusion equation.[10] The solution visualizes two elongated vortex cores periodically appearing at the absorption site due to absorption of one photon and streaming to opposite edges of the wire. Motion of vortices locally heats the superconductor; the cores become longer and eventually bridge the wire completely. The current corresponding to the appearance of the normal belt across the wire increases with the decrease in the photon energy. Like the quasistatic model, the time-dependent approach predicts the critical line in the phase diagram that demarcates regions with the current or wavelengths dependent detection efficiency and the region where the intrinsic detection efficiency reaches 100 per cent. Although photon excitation has been introduced differently in the quasistatic and the time–dependent approach, the effect of an absorbed photon was in both cases simplified. The quasistatic model considers uniform suppression of the order parameter over the width of the wire whereas the numerical approach uses a local electron heating in a circular spot with a wavelength independent diameter as an initial condition for numerical simulations. This may cause discrepancies with the experimental data since all measurements reported so far were done on wires much wider than the electron thermalization length.

Experimentally the spectral cut-off in the intrinsic detection efficiency was first reported one decade ago[11] and has been since then observed by several groups in detectors from different materials.[12,13,14] It has been shown that changing the wire thickness or film stoichiometry, e.g. in NbN, shifts the cut-off quantitatively according to the hot-spot model.[4,15] However, measured cut-off wavelengths differed noticeably from the values computed in the framework of this model. The problem most likely was that varying the thickness or stoichiometry influences practically all material parameters and hence increases uncertainty in quantitative comparison of the experimental data with the model.

In this paper we study the spectral cut-off in nanowires with different widths but made from the very same superconducting film which had well defined superconducting and metallic properties. We quantitatively compare our experimental results with available models of the critical state and show that the hot-spot model better describes experimental observations. In order to comprehend whether the discrepancy between the vortex-assisted scenario and the experimental data originates from the vortex scenario itself or from modeling the photon excitation, we also compare the dark count rates in our meanders with the predictions of the vortex model. The paper is organized as follows: In Section II known theoretical models are compared in more detail. We then describe the specimens and the experimental setup in Section III and the experimental results in Sections IV and V. The discussion and conclusion complete the paper.

## II. THEORETICAL MODELS

Here we present an overview of the existing models and their predicted cut-offs. In the original concept[1] it was supposed that when an optical photon is absorbed by an electron in a superconducting nanowire with a transport current, it creates a normal spot where the superconducting order parameter is suppressed. The current is then forced to flow around the spot. If in the sidewalks it exceeds the depairing current $I_{dep}$ the superconducting state will be destroyed locally and the superconductor will go to the resistive state. If one assumes uniform current distribution in the sidewalks, the value of the transport current $I_0$ at which this happens is easily found from the following obvious condition: $I_0/I_{dep} = 1-2R_N/w$ where $R_N$ is the radius of the normal spot and $w$ is the width of the wire. In this simplest model the radius is found by equalizing the superconducting condensation energy in the region with the size $\pi(R_N)^2 d$ ($d$ is the film thickness) with the energy of the incoming photon $hc/\lambda$. By combining the expressions above one may find the so called red boundary or cut-off wavelength $\lambda_0$ above which photon detection cannot occur at given values of the transport current $I$ and the wire width

$$\lambda_0 = \frac{8\varsigma}{\pi\, d\, N_0\, \Delta^2}\frac{h\,c}{w^2}\left[1-\frac{I}{I_{dep}}\right]^{-2} .\qquad(1)$$

The coefficient $0 < \varsigma < 1$ accounts for losses of the photon energy via electron-phonon interaction and phonon escape into the substrate that reduces the size of the normal spot. In Eq. (1) $\Delta$ is the equilibrium superconducting energy gap and $N_0$ is the density of states for normal electrons at the Fermi level.

This model was later refined to take into account non-homogeneous suppression of the order parameter due to diffusion of nonequilibrium quasiparticles. In Ref. 3 it was noticed that it is not necessary to destroy the superconductivity completely in the hot spot to have the resistive state. Indeed, suppression of the superconducting order parameter lowers the ability of the superconductor to carry superconducting current (the simplest example – suppression of $\Delta$ with increasing temperature which leads to a decrease in the depairing current). Authors of Ref. 3 noticed that if in the belt-region with the size roughly $\xi\, w\, d$ ($\xi$ is the superconducting coherence length) the mean concentration of superconducting electrons $n_S \approx N_0\, \Delta$ is reduced to the value $n_S' < n_S$ than the critical current will be locally lowered in comparison with the equilibrium depairing current to $I_0 = I_{dep}\, n_S'/\, n_S$. They further assumed uniform supervelocity across the wire and presented the hot spot as a two-dimensional circle with the exponentially decaying concentration of hot quasiparticles $n_n(r) \approx \exp(-r^2/4\, D\, \tau_{th})$ where $D$ is the diffusivity of normal electrons and $\tau_{th}$ is their thermalization time. The decrease in the number of superconducting electrons $(n_S - n_S')\,\xi\, w\, d$ in the belt region was then related to the increased number of quasiparticles meaning that creation of hot quasiparticles is possible at low temperatures only via destruction of superconducting electrons. Assuming that $\xi << (D\, \tau_{th})^{1/2}$ and $\tau_{th} << \tau_{eph}$ ($\tau_{eph}$ is the electron-phonon interaction time), they presented the number of quasiparticles as $\xi\, d \int_{-\infty}^{\infty} n_n(\,r\,)dr$ and finally obtained the cut-off wavelength

$$\lambda_0 = \frac{\varsigma}{d\, N_0 \Delta^2}\frac{h\,c}{w\sqrt{\pi\, D\, \tau_{th}}}\left[1-\frac{I}{I_{dep}}\right]^{-1}\qquad(2)$$

In order to explain the non-vanishing detection ability of superconducting nanowire single-photon detectors at $\lambda > \lambda_0$, which was observed in experiments, the authors of Ref. 16 suggested a vortex-assisted detection mechanism. The main idea is that the absorbed photon locally suppresses the superconducting order parameter (along with the model of Ref. 3) that favors the creation of thermally activated vortex-antivortex pairs around the absorption site. Such pairs are then unbound by the transport current. Motion of these vortices under Lorentz force heats the superconductor and destroys superconductivity.

Recently two models have appeared which also considered the vortex assisted mechanism of photon detection.[4,6,10] The approach of Ref. 6 implies homogeneous decrease of the superconducting condensation energy across the nanowire width by the absorbed photon. This suppresses the critical current and enhances locally the probability for single vortices to penetrate and cross the nanowire due to thermal activation. The authors argue that the energy barrier for activation of a vortex-antivortex pair is much larger than the barrier for a single vortex and hence the second process has to dominate. The single-vortex scenario should explain the finite but rapidly decreasing detection efficiency at $\lambda > \lambda_0$. The model predicts that the detection efficiency saturates at 100 per cent when the barrier for vortex crossing locally disappears. Formally this condition can be used to define the cut-off wavelength for particular values of the transport current and the wire width. Since the critical current $I_C$ in this model is defined as the current at which the barrier disappears, such definition of the cut-off coincides with the one from Ref. 3. The number of nonequilibrium quasiparticles reaches maximum at the time $\tau_{th}$ after the photon has been absorbed, therefore the largest uniform decrease in the superconducting condensation energy may occur in the wire area with the size $A = w^2$ if $w > (D\,\tau_{th})^{1/2}$ or $A = w\,(D\,\tau_{th})^{1/2}$ in the opposite case. Taking into account the dependence of the critical velocity of superconducting electrons on their concentration one finds the connection between the local critical current and the uniform concentration $I_0 = I_C\,(n_s'/n_s)^{3/2}$. Note that at $T \approx T_C$ the concentration $n_S \propto (1\text{-}T/T_c)$ and the superconducting condensation energy density $F \propto n_S{}^2 \propto (1\text{-}T/T_C)^2$ while the critical (depairing) current $I_{dep} \propto (1\text{-}T/T_C)^{3/2}$. Therefore the ratio of critical currents at different temperatures is proportional to $(n_s'/n_s)^{3/2}$. Associating the superconducting condensation energy density $F$ with $n_S$ and the energy gap $F = (n_S)^2/N_0 = N_0\,\Delta^2/2$, one finds the connection between the concentration and the photon energy

$$\left(\frac{n_s'}{n_s}\right)^2 = 1 - \frac{hc}{\lambda}\frac{2\varsigma}{N_0\Delta^2 A d}\,.\qquad(3)$$

Combining Eq. (3) with the expressions above we approximate the cut-off wavelength which follows from this model

$$\lambda_0 = \frac{2\varsigma}{d\,N_0\Delta^2}\frac{hc}{A}\left[1 - \left(\frac{I}{I_C}\right)^{4/3}\right]^{-1}\,.\qquad(4)$$

We note here that this simplified expression is only helpful for comparison with other models. To quantitatively describe the experimental data the exact result (Eq. (45), Ref. 6) has to be used.

A different vortex assisted detection mechanism was suggested in Ref. 10 for photon wavelengths $\lambda < \lambda_0$. The approach relies on the hot spot model in that it considers an initial heating of electrons in the spot with the radius $R_0 \approx (D\,\tau_{th})^{1/2}$, which evolves into a time dependent normal spot surrounded by an area with partially suppressed order parameter. The subsequent distributions of the electron temperature and the order parameter were found from the time-dependent Ginsburg-Landau equation and the heat diffusion equation which were coupled in space and time. The analytical extension of the model explicitly takes into account that the distributions of both the supervelocity and the current density are strongly non-uniform around the spot. To analytically define the size of the spot it was postulated that the strongest effect on the current redistribution is produced when the electron temperature in the spot center equals the superconducting transition temperature $T_C$. This provides the following connection between the photon wavelength and the spot radius $\pi\,d\,R^2\,c_V\,(T_C\text{-}T) = hc/\lambda$ where $c_V$ is the electron specific heat capacity and $T$ is the bath temperature. When the Supervelocity reaches its critical value near the edge of the spot, the conditions are created for nucleation of a vortex and antivortex on opposite sides of the spot and their consequent motion across the wire (in more general case of a spot with partially suppressed order parameter the vortex-antivortex pair is nucleated inside the spot and vortices can leave the spot when the supervelocity near the edge reaches its critical value). This happens if the transport current is larger than some certain value $I_0$, which depends on the instant radius of the spot $R$, the wire width $w$ and the degree of suppression of the order parameter in the spot. In the framework of the London model it was found that this current value is

$$\frac{I_0}{I_{dep}} = \left[1 - \left(\frac{2R}{w}\right)^2 \frac{1-\gamma^2}{1+\gamma^2}\right] \bigg/ \left[1 + \frac{R}{R+\xi} \frac{1-\gamma^2}{1+\gamma^2}\right], \qquad (5)$$

where $\gamma$ defines the ratio of the superconducting order parameter inside and outside the spot and can be expressed via the respective concentrations of superconducting electrons as $\gamma = (n_S{}'/n_S)^{1/2}$. Note that when $R$ approaches $w/2$ and simultaneously $w >> \xi$ and $\gamma = 0$ (which corresponds to the normal spot) $I_0$ in Eq. (5) coincides with $I_0$ found in the framework of the normal-spot model (see expression before Eq. (1)). This occurs because in this limiting case one can neglect the non-uniformity in the current density distribution in the sidewalks around the normal spot. An analytical solution for $\gamma$ was found in the form

$$\gamma = exp\left(-\frac{h\,c}{\lambda} \frac{\varsigma}{c_v\,d\,4\pi\xi(0)^2 T_c} ln\left(-\frac{h\,c}{\lambda} \frac{\varsigma}{c_v\,d\,4\pi\xi(0)^2 T_c}\right)\right), \qquad (6)$$

where $\xi(0)$ is the Ginsburg-Landau coherence length at zero temperature. Combining Eq. (6) (in the simplest case of the normal spot with $\gamma=0$) with the definition of $R$ one can obtain the cut-off wavelength as the solution of the following cubic equation

$$\left(\frac{2\xi}{w}\right)^2 u^3 + \left(\frac{2\xi}{w}\right)^2 u^2 + \left(2\frac{I}{I_{dep}} - 1\right)u + \left(\frac{I}{I_{dep}} - 1\right) = 0, \qquad (7)$$

where the variable $u = R/\xi$ contains $\lambda_0$.

It follows from Eq. (7) that with the increase in the relative transport current $I/I_{dep}$ the dependence $\lambda_0(w)$ gradually changes from $\lambda_0 \propto w^{-2}$ at $I/I_{dep} << 1$ to the width independent cut-off wavelength. At small currents only those photons are detected which create the spot comparable with the wire width and the dependence $\lambda_0(w) \propto w^{-2}$ coincides with that in Eqs. (1) and (4). At currents close to the depairing current even spots with $R_0 << w$ ($w$ should still be smaller than the Pearl length) make the superconducting state unstable.[10]

### III.  STUDIED MEANDERS AND EXPERIMENTAL SETUP

In this work meanders made from thin films of TaN and NbN were investigated. The films were deposited on sapphire and silicon substrates respectively by dc reactive magnetron sputtering. To prevent back-reflection substrates were polished from one side only. NbN films were grown on an additional 250 nm thick buffer layer of silicon oxide in order to increase photon absorption at near-infrared wavelengths. The films were then structured into meanders by electron beam lithography and reactive ion etching leading to a photon active area of approximately 4x4 μm² with different line widths. The spacing between lines was kept almost constant in TaN meanders while it varied in meanders from NbN. The ends of the meander line were connected to a coplanar transmission line, which had 50 Ohm impedance in order to match the microwave amplifier and cable impedances and to prevent latching in the resistive state. For further information on the fabrication process see Refs. 17 and 18. In Fig. 1 a scanning electron microscope (SEM) image of a typical TaN meander (a) and an atomic force microscopy (AFM) image of one of the studied NbN meanders (b) is displayed.

It has been shown theoretically[19] and confirmed experimentally[20,21] that the critical current is to a certain extent limited by current crowding near the bends. Therefore special attention was paid to structure the bends in the meanders as smooth and round as possible in order to increase the experimental critical current towards the depairing value.

Geometrical meander characteristics like the line widths, active area and spacing where determined by scanning electron microscopy. Furthermore the sheet resistance of the specimen was determined by the 20K normal-state resistance together with the wire widths and lengths. Important meander parameters that are used in this paper are summarized in Table I.

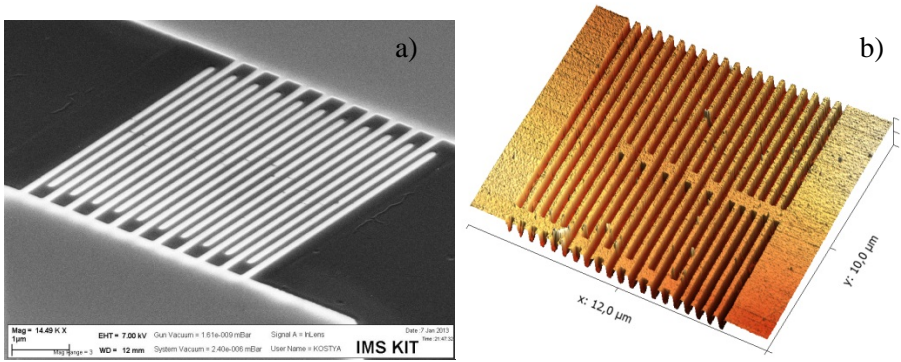

FIG. 1. (a) SEM image of a mask from photoresist which defines the shape of TaN meander. The black portion corresponds to the meander line. (b) AFM image of a NbN meander used in this study.

TABLE I. Characteristics of the TaN and NbN meanders which were used to measure detection efficiencies. The depairing critical current was calculated according to Bardeen (Ref. 22) for the extremely dirty limit with a correction by Kuprijanov-Lukichev (Ref. 23).

| Sample No. | Thickness of the nanowire | Width of the nanowire | Transition temperature | Square resistance | Experimental critical current @ 4.5 K | Depairing critical current @ 4.5 K |
|---|---|---|---|---|---|---|
| | (nm) | (nm) | (K) | (Ω) | (μA) | (μA) |
| TaN1 | 4.0 | 73 | 8.6 | 386 | 10.0 | 22.2 |
| TaN2 | 4.0 | 92 | 8.7 | 376 | 15.0 | 29.6 |
| TaN3 | 4.0 | 110 | 8.9 | 407 | 17.9 | 35.1 |
| TaN4 | 4.0 | 112 | 9.1 | 396 | 19.8 | 38.4 |
| TaN5 | 4.0 | 133 | 8.9 | 414 | 20.3 | 41.7 |
| TaN6 | 4.0 | 146 | 9.6 | 517 | 25.4 | 42.1 |
| TaN7 | 4.0 | 179 | 9.6 | 559 | 28.1 | 47.7 |

| | | | | | |
|---|---|---|---|---|---|
| TaN8 | 4.0 | 220 | 9.2 | 470 | 32.1 | 65.9 |
| TaN9 | 4.0 | 243 | 8.92 | 433 | 33.0 | 72.9 |
| NbN1 | 3.6 | 122 | 9.0 | 586 | 36.6 | 59.7 |
| NbN2 | 3.6 | 130 | 9.2 | 580 | 41.2 | 61.0 |
| NbN3 | 3.6 | 156 | 10.2 | 878 | 36.1 | 48.4 |
| NbN4 | 3.6 | 178 | 10.0 | 683 | 35.8 | 70.9 |
| NbN5 | 4.8 | 85 | 8.5 | 569 | 32.0 | 48.1 |
| NbN6 | 4.8 | 98 | 8.7 | 453 | 22.0 | 41.3 |
| NbN7 | 4.8 | 130 | 9.4 | 424 | 44.0 | 68.9 |

Fig. 2 shows the experimental open-beam setup that was used to obtain the intrinsic detection efficiency. The substrate with the meander is fixed to a copper holder to ensure good thermal stability. This holder is mounted to a cold plate inside the vacuum chamber of a $^4$He bath cryostat, thus temperatures of about 5 K at the meander can be reached with this setup. Additionally an electric circuit working as a bias tee is installed on the copper holder where the detector is wire bonded to. Biasing of the meander was done by a tunable low noise self-made dc voltage source. Voltage signals generated by the meander due to absorption of single photons are guided out of the cryostat by a coaxial cable, amplified by two microwave amplifiers with 28 db and 20 db and finally recorded by a 200 MHz pulse counter.

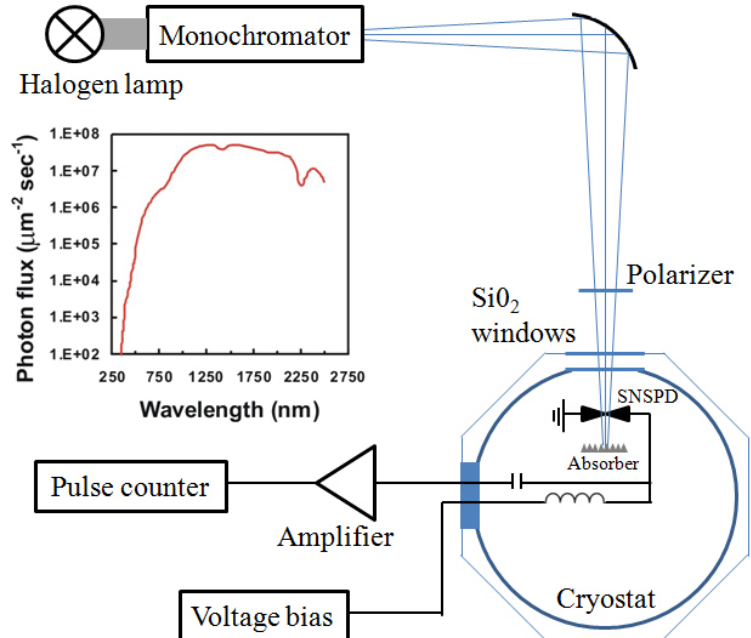

FIG. 2. Schematics of the open-beam experimental setup. The plot shows the photon flux available at the position of the detector as a function of the wavelength. This curve was used to normalize measured photon count rates and to extract the meander detection efficiency.

In order to illuminate the sample a halogen lamp is used as a light source. The emitted incandescent light coming from the lamp is passed through a prism monochromator where wavelengths from 300-2500 nm can be selected. The monochromatic beam is then expanded to a few millimeters in diameter to assure a homogenous light field for illumination of the few μm² sized meander. Since the meander is most sensible to light polarization parallel to its lines, a polarizer is placed in the optical path. The beam then passes through 2 parallel SiO₂ windows and a 2 mm aperture inside the cryostat and hits the front surface of the detector at right angles. Transmitted light is absorbed several millimeters behind the rear side of the detector to ensure that no light is scattered back into the meander. To obtain the intrinsic detection efficiency (*IDE*) from the experimentally measured count rates it is essential to know the photon flux (*PF*) at the exact position of the meander. Therefore we installed certified photodiodes at the meander position inside the cryostat and measured the photon flux as a function of the wavelength (inset in Fig. 2). The detection efficiency at each wavelength DE(λ) was defined as the difference between the photon count rate (*PCR*) and the dark count rate (*DCR*) related to the photon flux (*PF*) incident upon the meander area. *IDE* was calculated using the following relation: *IDE = (PCR-DCR)/(A PF Abs)*, where *A*, denotes the meander area and *Abs* the absorbance of the meander structure. The wavelength-dependent absorbance per unit area of a meander structure was numerically computed with the account of the filling factor, wire width and the square resistance of the film. Details of the numerically computed absorbance can be found in Ref. 24.

## IV. EXPERIMENTAL RESULTS: PHOTON COUNTS

Spectral measurements of the *IDE* were performed on all detectors listed in Table I. For each detector spectra were recorded at three bias currents of 0.8, 0.87 and 0.95 times the experimental critical current. In terms of the depairing critical current the ratios varied between 0.5 and 0.7. Fig. 3 exemplarily depicts one of the *IDE* spectra obtained with a TaN (squares) and with a NbN (circles) meander with 112 nm and 130 nm line width respectively. Due to the 250 nm thick SiO$_2$ buffer layer in NbN detectors the detection efficiency quickly oscillates at small wavelengths when they are comparable to the layer thickness. These oscillations are caused by oscillations in the meander absorbance which can be numerically computed solving a multilayer system. The inset of Fig. 3 shows the computed *DE* within a multilayer system compared to the experimentally obtained detecting efficiency. Since the positions and the strengths of the maxima in the computed absorbance are strongly affected by the thickness of the buffer layer and the absorbance at the interfaces, normalizing the detection efficiency with the absorbance increases the uncertainty. Furthermore the light intensity provided by the monochromator at wavelengths smaller than 450 nm decreases very fast towards zero thus increasing the relative error in the *DE* in this range up to 35 %. Both these factors cause the peak at 400 nm in the *IDE* of NbN meanders. Overall, however, *IDE* spectra look quite similar for detectors from NbN and TaN. Coming from small wavelengths the *IDE* is constant at 100 % within the experimental accuracy. With decreasing photon energy the *IDE* transitions into a steep slope which matches a power law. Comparison of two materials confirms earlier observation[25] that for close relative currents NbN is less effective at larger wavelength than TaN. Partly this drawback is relaxed by the difference in the relative operation temperatures and the temperature dependence of the cut-off wavelength.[26]

### A. Experimental cut-off wavelengths

In order to extract the cut-off wavelength from the experimental data formal fitting of the measured IDE spectra with the following expression was used:

$$IDE(\lambda) = \frac{IDE_0}{\left[1 + (\lambda / \lambda_0)^{n/2}\right]^2}$$

(8)

$IDE_0$ denotes the detection efficiency at the plateau of the spectrum at small wavelengths whereas the denominator stands for the power law like decrease of the *IDE* at larger wavelengths. $\lambda_0$ is the cut-off wavelength. This analytical approximation was proven to best describe the *IDE* spectra of TaN at all experimental conditions. However we would like to stress, that this analytical fit has no physical meaning and was introduced in order to formalize the extraction of the cut-off wavelength from the experimental data.

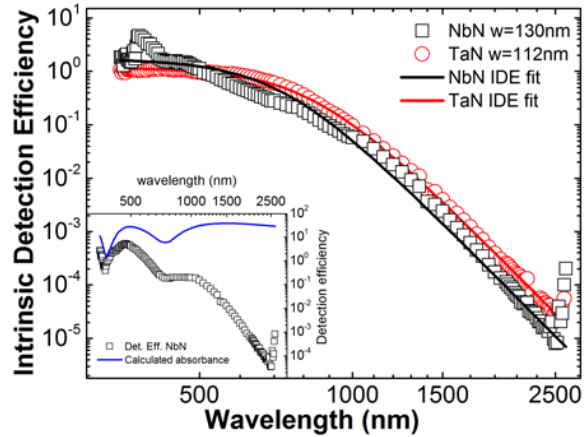

FIG. 3. Intrinsic detection efficiency of a TaN and a NbN meander with a respective wire width of 112 nm and 130 nm at bias currents of 0.47 and 0.52 of the depairing current. Solid lines show the fit of Eq. (8) to the data. The inset shows the oscillating detection efficiency of the NbN meander (squares) due to an additional SiO$_2$ buffer layer and the calculated absorbance of this multilayer system (solid line) in per cent.

In the following the three theoretical approaches, which were introduced in the second section of this paper, are used to fit the experimental data. Formulas are explicitly expressed in SI units. All expressions were converted in a form only depending on parameters of the meander that can be directly measured. The parameters that were used in the fitting procedure are summarized in Table II.

TABLE II. Parameters of TaN and NbN meanders that were used to calculate the different theoretic width dependencies of the cut-off wavelength.

| | Average square resistance | Multiplication efficiency | Diffusion coefficient | Electron thermalization time | Coherence length at 4.5 K | Energy gap at 4,5K | Thickness of the nanowire | Ratio of the bias current to the crit. dep. current |
|---|---|---|---|---|---|---|---|---|
| | ($\Omega$) | | (cm$^2$s$^{-1}$) | (ps) | (nm) | (meV) | (nm) | |
| TaN | 450 | 0.38 | 0.6[a] | 7 | 7.0 | 1.27 | 4.0 | 0.50 |
| NbN | 600 | 0.43 | 0.5 | 7 | 5.8 | 1.77 | 3.6 | 0.62 |
| NbN | 482 | 0.43 | 0.5 | 7 | 5.8 | 1.81 | 4.8 | 0.70 |

[a] This diffusion coefficient for TaN was taken from Ref. 25; For all specimens we used the same electron thermalization times from Ref. 27.

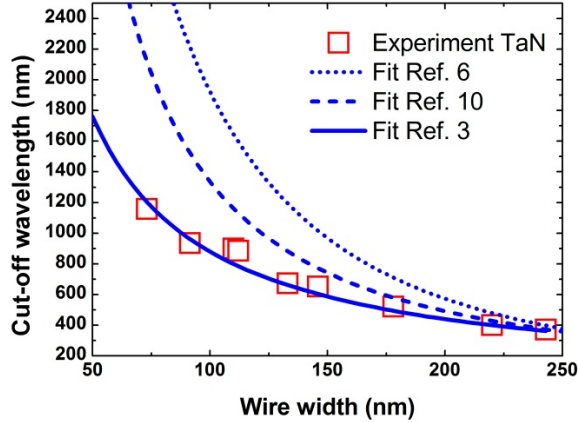

FIG. 4. Cut-off wavelength of the TaN set (squares) versus the meander wire width compared to three theoretical models. The dashed and dotted lines corresponds to the vortex assisted detection models and the solid line to the hot spot model.

Fig. 4 shows the cut-off wavelengths (squares) of the TaN specimens with meander line widths varying from 65 nm to 250 nm and the three fits made with different models. The dotted line corresponds to the vortex-assisted hot-belt detection model;[6] the dashed line shows the fit with the vortex-assisted hot-spot model[10] and the solid line corresponds to the non-homogeneous hot spot model.[3] Although it does not visibly change the plots, we also accounted for the degradation of the wire edges[28] shifting all theoretical curves by 5 nm towards smaller wire widths. The quantum efficiency or quantum yield $\varsigma$ defines the portion of the energy of the absorbed photon which is transferred to quasiparticles at the time of maximum local reduction of the number of Cooper pairs (energy gap). This portion is further denoted as effective photon energy. To simplify comparison in Fig. 4 $\varsigma$ was adjusted in a way that all curves intersect at a wire width of 250 nm and hit at this point the experimental cut-off wavelength. Varying $\varsigma$ only shifts the curves vertically and does not change their forms. The experimentally obtained cut-off wavelength clearly decreases with increasing the wire width which is qualitatively predicted by all three theoretic models. However the non-homogeneous hot-spot model most precisely describes the experimental data.

### B. Vortex assisted hot-belt model: quasistatic approach

We adopt the assumption of Ref. 6 that the cloud of nonequilibrium quasiparticles has uniform density and spans the entire width $w$ of the wire with the thickness $d$. Since for TaN $w > (D\,\tau_{th})^{1/2}$ the effective photon energy is homogeneously distributed in the volume $(w^2 d)$ where it reduces the superconducting condensation energy $F$. With the obvious relation $F = \varepsilon_0\,(w/\xi)^2/(8\,\pi)$ where $\varepsilon_0$ is the vortex energy scale (Ref. 6, Eq. (8)) and $\xi$ is the coherence length, we find the reduced value of the factor $\nu = \varepsilon_0/(k_B\,T)$ in the cloud

$$\nu_h = \nu - 4\pi\varsigma\,\frac{h\,c}{\xi\,k_B T}\left(\frac{w}{\xi}\right)^{-3}\left(\frac{\lambda}{w}\right)^{-1}. \qquad (9)$$

We further identify our experimental critical current $I_C$ with the current at which the vortex barrier vanishes (Eq. (2)) and formally define the cut-off wavelength $\lambda_0$ via Eq. (45) of Ref. 6) as the wavelength at which the term under the exponent equals one. Then for the fixed relative bias current we varied the wavelength looking for the wire width that kept the exponential term constant. The relation between $\lambda_0$ and $w$ is shown in Fig. 4 by the dotted line. It was obtained for the relative bias current $I_B$=0.9 $I_C$, $\xi = 7$ nm, $\varsigma = 0.28$, and the value of the Pearl length $\Lambda$=118 µm, which was calculated as $\Lambda = 2\,\hbar\,R_S/(\pi\,\mu_0\,\Delta)$. Here $\Delta$ is the superconducting energy gap at the operation temperature and $R_S$ the square resistance. We use an effective relaxation time of the hot-spot of 10 ps (see Eq. (43) in Ref. 6) and the actual parameters of our wires (Table II). The vertical position of the curve is very sensitive to the bias current and the quantum efficiency while other parameters only slightly affect its position and curvature. Compared to the experimental data, the model predicts a much larger rate of the increase in $\lambda_0$ with the decrease in the wire width.

### C.   Vortex assisted photon counting model: Time-dependent Ginsburg Landau approach

For our wavelength range (0.3 to 2.5 µm) and the material parameters of TaN the factor $\gamma^2$ (Eq. (6)) is less than 0.01 hence allowing us to use Eq. (7). We identify our experimental transport current with the current $I_0$ in Eq. (5) and solve Eq. (7) numerically for each ratio $\xi/w$ to obtain the cut-off wavelength. Following the definitions in Section II one can present the variable u in the form $u = R/\xi = $ (h c $\varsigma/(\pi\,\lambda_0\,c_V\,d\,T_C\,\xi(0))^{1/2}$ with material parameters from Table I and Table II and the specific electron heat capacity 2.04 mJ/cm³, which was computed as $c_V = \pi^2\,k_B^2\,T_C^2/(3\,e^2\,R_S\,d\,D)$ where e is the electron charge, we find for TaN $u = (33.4\,\varsigma/\lambda_0[\mu m])^{1/2}$. The resulting dependence $\lambda_0(w)$ which was obtained for $\varsigma = 0.28$ and the relative current $I/I_{dep} = 0.5$ is shown by the dashed line in Fig. 4. The combination of the time-dependent GL approach and the London model better describes experimental data for larger wire widths but overestimates the cut-off wavelengths for narrow wires.

### D.   Non-homogeneous hot-spot model

To compare our experimental results with this model[3] we modified Eq. (2) taking into account that in TaN $\tau_{th} \leq \tau_{eph}$ and using the square resistance instead of the electronic density of states.[4] The cut-off wavelength or the maximum photon wavelength that is needed to form a resistive barrier over the entire wire width reads:

$$\lambda_0 = \varsigma \frac{4\,R_S\,e^2}{3\,\pi^{1/2}\,\Delta^2}\,\frac{h\,c}{w}\,\sqrt{\frac{D}{\tau_{th}}}\left[1 - \frac{I}{I_{dep}}\right]^{-1} \qquad (10)$$

Since the square resistances and (hence the depairing critical current) of the samples differ in TaN by about 5% and in NbN up to 15% most likely due to variations in the thickness of the meander lines it is clear that one single plot for the theoretical cut-off wavelength vs. line width cannot account for all the measured data points at the same time. Therefore an averaged square resistance was calculated as a compromise to fit best to the data points. The hot spot model dependence of the cut-off wavelength on the wire width is shown in Fig. 4 by the solid line. The best fit to the data points was achieved with an efficiency of $\varsigma$=0.38. Overall the experimental data can be described best by the non-homogeneous hot spot model although it does not account for the steep decrease of the IDE on the long wavelength side of the cut-off.

We did not attempt to fit the experimental data for NbN with all available models since the range of the wire width is much smaller in this case and the variation in the square resistance is relatively large making the comparison not very instructive. Instead we fit the data for both materials with the non-homogeneous hot-spot model to additionally check its applicability. Fig. 5 shows the cut-off wavelength versus the reciprocal width of the wire. This representation is easily accessible and shows the data points for different materials and the model fit as straight lines. The main contribution to the error bars comes from the determination of the cut-off wavelength using Eq. (8). We find a relative error of approximately 7% for all data points.

Despite of the variation of the square resistance in our NbN samples it can be seen that with increasing wire thickness the cut-off wavelength decreases as predicted by the hot spot model. We find $\varsigma=0.43$ as the best fit for NbN and $\varsigma=0.38$ for TaN. Similar efficiency for NbN was already reported by Hofherr et al..[4] Experimental data for TaN and NbN ($d$=4.8) meanders can be fairly well described by the hot-spot model whereas the fitting line for another NbN ($d$=3.6) batch seems to be systematical shifted to higher cut-off values by about 10%. This discrepancy can only be explained by poor definition of either the transition temperature or the sheet resistance, since all other model parameters are the same for both NbN series.

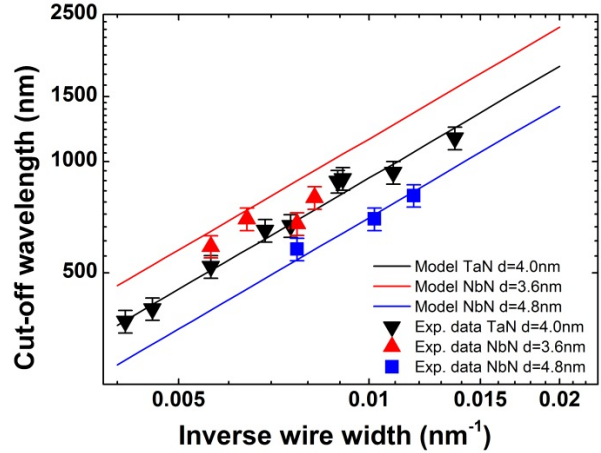

FIG. 5. Cut-off wavelength of the TaN and NbN meanders plotted against the inverse wire width. The solid lines represent the hot spot model (Eq. (10)) calculated with the averaged square resistance of the particular sample set. The error bars result from the extraction procedure of $\lambda_0$.

## V. EXPERIMENTAL RESULTS: DARK COUNTS

Since the photon counts at low photon energies are likely just dark counts in the portion of the line where the energy gap has been locally reduced after photon absorption,[4,5,6,29] and since the quasistatic model of the vortex-assisted photon detection[6] is just an extension of the dark count model,[30] we attempt to examine whether the dark count rate (DCR) alone in our meanders can be described with the quasistatic vortex model of Ref. 30. Dark count rates as functions of the bias current were measured at 4.2 K for NbN meanders with different wire widths. Fig. 6 (a) displays the results for seven different wire widths at a temperature of 4.2 K. It is noticeable that the applicable bias current increases with the wire width.

TABLE III. Characteristics of the NbN meanders presented in Fig. 6.
The set has a nominal thickness of d=3.6 nm.

| Wire width | Square resistance | Critical current for 4.2K | Transition temperature |
|---|---|---|---|
| ( nm ) | ($\Omega$) | ($\mu$A) | (K) |
| 94 | 757 | 8.6 | 8.0 |
| 108 | 956 | 14.6 | 8.5 |
| 118 | 889 | 16.8 | 9.1 |
| 124 | 927 | 19.8 | 8.8 |
| 156 | 546 | 40.4 | 10.2 |
| 169 | 774 | 39.3 | 9.9 |
| 182 | 683 | 48.8 | 10.3 |

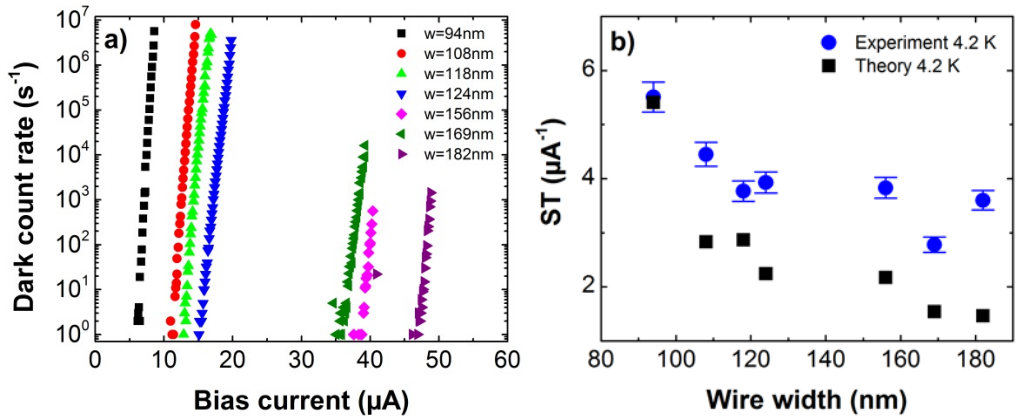

FIG. 6. (a) Dark count rates (DCR) as a function of the bias current of seven meanders with different wire widths recorded at 4.2K. (b) Steepness (ST) of the DCR in logarithmic scale versus the actual wire width of the meander. Circular symbols represent the experimental data whereas the squared symbols stand for theoretical modeling according to Ref. 30. The slope extraction and the error bars are explained in detail in the text.

For comparison of the experimental DCR with the theoretical rates, Eq. (51) of Ref. 30 was reworked to include only directly measurable physical quantities. We than computed for each specimen from Table III the steepness of the current dependence of DCR as follows:

$$ST = \frac{d}{dI}\left(Ln\left(DCR(I)\right)\right) = \frac{\left(\frac{\Phi_0}{\pi v k_B T}\right)^2 I}{1+\left(\frac{\Phi_0}{\pi v k_B T}\right)^2 I^2} + \frac{\Phi_0}{\pi k_B T} arctan\left[\left(\frac{\Phi_0}{\pi v k_B T} I\right)^{-1}\right] \quad (11)$$

$$\nu = \frac{\mu^2 \varepsilon_0}{k_B T} = \frac{\mu^2 \Phi_0^2 \Delta(0)}{4\hbar R_S k_B T}\left[1-\left(\frac{T}{T_C}\right)^2\right]^1 \left[1+\left(\frac{T}{T_C}\right)^{\frac{3}{2}}\right]^{\frac{1}{2}} \quad (12)$$

Physical parameters that enter the steepness (ST) in Eq. (11), which is the total derivative of the DCR in logarithmic form, are summarized in Table III. The critical temperature for this series of NbN meanders is $T_C = 10.7$ K and the value of the energy gap $\Delta(0) = 2.02\ k_B T_C$ was taken from Ref 15. Temperature dependence in ν (Eq. (12)) stems from the GL temperature dependence of the penetration depth and differs from the BCS temperature dependence of the energy gap by less than 3%. The factor μ²=1-κ², where κ stands for the order parameter suppression due to the bias current and varies from 0 at T=0 K to $3^{-1/2}$ at $T = T_C$. Since the measured critical current is about 0.5 from $I_{dep}$ (see Table II) we neglect the dependence μ($I$) and choose μ to be unity. In Fig. 6 (b) ST is displayed over the wire width $w$ of the meanders. From our standpoint it is the most expressive and accessible representation. Slope extraction of the experimental data was done by fitting the expression $y=exp(a\ x-b)$, where $a$ is the slope and $b$ the y-intercept, to the five largest measured current values of each meander in Fig. 6 (a). We estimated an extraction error of 5% to originate from that procedure. Concerning the theoretic model the above derived expression for DCR was taken in logarithmic form and linearized for each measurement at a current $I = 0.98\ I_C$ (see Table III) by a first order Taylor series. First order expansion is sufficient since the function in the range of interest is almost nearly linear.

The strong variation of square resistances in the series (Table III)) might be explained by fabrication processes. The thickness of the film may vary by about 10% from the edges to the middle of the wafer. As already reported[24] the square resistance varies non-linearly with the reciprocal thickness concerning films less than 6 nm thick. Given the case, that the measured meanders are from different places on the wafer and hence have a variation of 10% it would cause a large variation of the square resistance from about 600-1000 Ω. This approximation matches the measured square resistances for these samples (see Table III) very well.

## VI. DISCUSSION AND CONCLUSION

Although overall, the absolute values of the steepness in the DCR($I$) curves and its variation with the wire width are both fairly well reproduced by the theory of Ref. 30, at large widths there is a systematic difference between the computed steepness and the experimental values. This might be due to the different effect of the boundary conditions at the wire edges for the vortex barrier in wires with different widths.[7,9] Since the photon counting model[6] is an extension of the DCR model for photon excitation, we believe that the discrepancy, which apparently appears between the experimental cut-off wavelengths and the cut-offs computed in the framework of the extended model, is most likely due to the way how the photon excitation has been introduced. Indeed, accounting for a non-uniform excitation in the form of a spot with suppressed superconductivity in the time-dependent GL approach[10] results in a closer match of the model fit and the experimental data. Even better agreement is achieved with the non-homogeneous hot-spot model[3] where the gradual Gaussian distribution of nonequilibrium quasiparticles was explicitly taken into account. We argue below that a modification of the model of Ref. 10 that includes a more realistic excitation scheme drastically improves the agreement with the experimental data. If one formally uses a slightly larger value of the relative current $I/I_{dep}$ =0.55 and smaller value of ς = 0.11, the vortex assisted GL hot-spot model would fit the experimental data much better resembling the results of the non-homogeneous hot-spot model.[3] The following consideration explains why these parameters are more realistic. In Ref. 10 the radius of the hot spot $R$ was defined from the condition that, when the temperature in the center of the hot spot reduces to $T_C$, the temperature of the quasiparticles at $r = R$ equals $T(R) = T + (T_C-T)/e$ (where e=2.71). Using a little bit different criterion $T(R) = T + (T_C-T)/e^{1/2}$ reduces the effective radius of the hot spot by the factor $2^{1/2}$ and hence decreases the best fit value of the effectiveness by the same factor. Eq. (5) was obtained under the assumption that the order parameter exhibits a step at $r = R$, while the original numerical model outputs a smeared profile of the order parameter (or the density of superconducting electrons) similar to the one of the non-homogeneous hot spot model. The order parameter has a minimum in the center of the spot and grows exponentially at larger distances following variations of the local electron temperature. Additional area with the reduced order parameter will additionally decrease $I_0$ as compared to the step-like profile. This will result in a factor close but smaller than the one in the relative current in Eq. (7) which will make the best fitting value of the relative current closer to experimental value $I / I_{dep} = 0.5$.

In summary, we compared the spectral cut-off in the intrinsic detection efficiency of TaN and NbN nanowire meanders with different widths with the available models of photon counting and showed that the non-homogeneous hot-spot model best describes the variation of the cut-off wavelength with the wire width. We also found that the vortex model without photon excitation explains relatively well experimental data on dark counts in similar meanders. We argued that in any model of photon counting implementing the real Gaussian-like distribution of quasiparticle excitations improves the agreement between the experimentally measured cut-off wavelengths and the model predictions. Relying on this observation, we suggested that the experimental cut-off in the detection efficiency is determined by the lack of the current-carrying ability of the superconducting condensate. Furthermore, relatively good agreement between experimental data and different theories justifies that in a nanowire it is the proximity to the depairing current that governs the efficiency of photon detection rather than the proximity to the experimental critical current. We suggest that it should be possible to extend the detection efficiency to larger wavelength by improving the wire uniformity and increasing the experimental critical current up to a value closer to the depairing critical current.

## ACKNOWLEDGMENT

R. L. acknowledges support by the Helmholtz Research School on Security Technologies. Y. K., A. T., A. K., and G. G. acknowledge support by the RFBR grant 12-02-31841 and by the FPS "Scientific and scientific-pedagogical personnel of innovative Russia for years 2009-2013" grant No 14.B37.21.1631.